\documentclass[aip,amsmath,amssymb,reprint,groupedaddress,floatfix]{revtex4-1}
\usepackage[utf8]{inputenc}
\usepackage[T1]{fontenc}

\usepackage{graphicx}
\usepackage{floatrow}
\usepackage[caption=false,font={large}]{subfig}
\usepackage{bbm}

\usepackage{amsmath}
\usepackage{amssymb}
\usepackage{color}
\usepackage{times}
\usepackage{float}
\usepackage[sort&compress]{natbib}

\usepackage[colorinlistoftodos,prependcaption,textsize=small]{todonotes}

\usepackage{hyperref}

\hypersetup{
  colorlinks=true,
citecolor=black,
linkcolor=black,
urlcolor=black
  }

\begin{document}
\preprint{AIP/123-QED}

\title{Interplay of noise induced stability and stochastic resetting}
\author{Karol Capa{\l}a}
\email{karol@th.if.uj.edu.pl}

\author{Bart{\l}omiej Dybiec}
\email{bartlomiej.dybiec@uj.edu.pl}

\author{Ewa Gudowska-Nowak}
\email{ewa.gudowska-nowak@uj.edu.pl}

\affiliation{Institute of Theoretical Physics, Department of Statistical Physics, Jagiellonian University, \L{}ojasiewicza 11, 30-348 Krak\'ow, Poland}


\date{\today}

\begin{abstract}
Stochastic resetting and noise-enhanced stability are two phenomena which can affect the lifetime and relaxation of nonequilibrium states.
They can be considered as measures of controlling the efficiency of the completion process when a stochastic system has to reach a desired state.
Here, we study interaction of random (Poissonian) resetting and stochastic dynamics in unstable potentials.
Unlike noise-induced stability which increases the relaxation time, the stochastic resetting may eliminate winding trajectories contributing to the lifetime and accelerate the escape kinetics from unstable states. In the paper we present a framework to analyze compromises between the two contrasting phenomena in a noise-driven kinetics subject to random restarts.

\end{abstract}


%
\maketitle

\setlength{\tabcolsep}{0pt}

%
%
\textbf{Ubiquity of observed natural nonequilibrium systems acting under influence of noises attracted much attention on theoretical and experimental studies of dynamical stochastic systems and noise-induced phenomena.
Stochastic resetting and noise-enhanced stability are different stochastic protocols which can be used to control the lifetime of states or efficiency of processes containing random components.
On the one hand, the stochastic resetting can be used to eliminate subotimal trajectories and in turn it can accelerate the exit kinetics or increase the search efficiency.
On the other hand, the effect of noise-enhanced stability owes its effectiveness to the presence of meandering trajectories, as the action of noise could induce emergence of very long trajectories.
Here, we show how the stochastic resetting can counterbalance the action of noise which induces the noise-enhanced stability.
As a result of the stochastic resetting, the system can be moved out of the metastable state and shifted to the desired state more easily.}

%
%
\section{Introduction and model}

Nowadays, it is well known and widely accepted that the noise in dynamical systems is not always detrimental, but it can also play a beneficial role \cite{hanggi1990,anishchenko1999,gammaitoni2009}.
The action of random noise underlines the occurrence of various noise-induced phenomena such as stochastic resonance \cite{mcnamara1989,gammaitoni1998},  resonant (and stochastic resonant) activation \cite{devoret1984,doering1992} or ratcheting effects \cite{magnasco1993,reimann2002}.
Moreover, the appraisal of the role and importance of random forces
has been not only studied theoretically, but recorded in many real-life situations and biological setups \cite{russell1999use,simonotto1997visual}.

The noise stabilizing effect (noise-enhanced stability, NES) \cite{agudov1998noise,agudov1999decay,agudov2001} appears for stochastic diffusion in potential profiles with metastable states.
A signature of NES in static potentials is the non-monotonic behavior of the average escape time and prolonged lifetime of metastable or unstable states observed at an optimal intensity of noise \cite{fiasconaro2003,dubkov2020statistics}. In a metastable fluctuating potential, NES has been observed regardless of the unstable initial position of the Brownian particle \cite{agudov2001}. The appearance of the effect has been reported in physical systems like overdamped and underdamped Josephson junctions \cite{Pan2009}, chemical Belousov-Zhabotinsky and Michaelis-Menten reactions \cite{Yoshi2008,fiasconaro2006co}, ecological systems \cite{Zeng2013} and cancer growth dynamics \cite{Marcinek}.

Another process that can be used to control the decay of states is stochastic resetting \cite{evans2011diffusion,evans2020stochastic} referring to situations where the dynamics of a system is stopped and started over.
Starting from anew can either increase or decrease the average time needed to hit the target or hinder or accelerate the course of the chemical reaction.
On the one hand, stochastic restarting can eliminate long trajectories, preventing the exploration of distant parts of the space.
On the other hand, by preventing the exploration of distant points, it can decrease chances of finding distant targets.
Within the model explored in this paper, the stochastic resetting can be understood as the external manipulation on the system that brings the Brownian particle (or, e.g., the concentration of chemical reactants) to the initial state.

Our presentation begins with a general discussion of the effect of noise-enhanced stability and its properties.
Subsequently, we discuss the NES phenomenon with the relation to a model of tumor growth and control \cite{fiasconaro2006co}.

\begin{figure}[!h]
    \centering
    \includegraphics[angle=0,width=0.75\columnwidth]{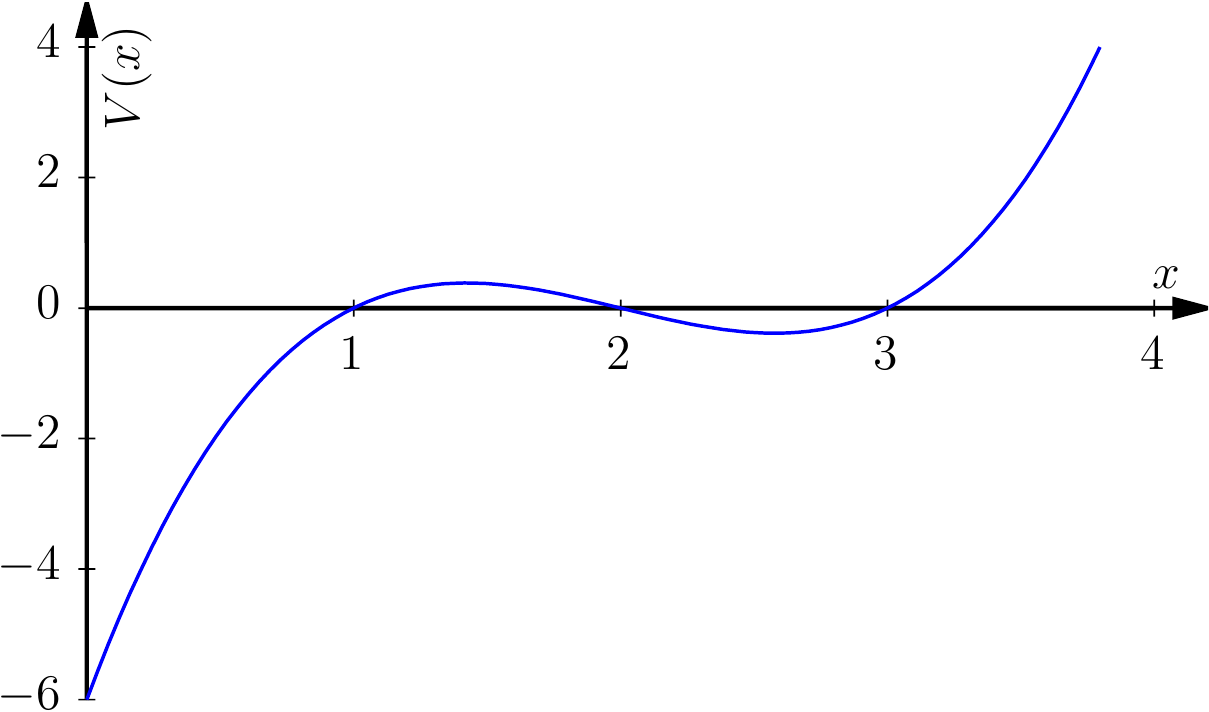}
    \caption{The cubic potential given by Eq.~(\ref{eq:potential}) used in the study of noise-enhanced stability under stochastic resetting.
    The top of the potential barrier is located at $x=2-1/\sqrt{3} \approx 1.42$, while a local stable state is associated with the minimum of the potential located at $x=2+1/\sqrt{3} \approx 2.58$.
    }
    \label{fig:potential}
\end{figure}

We study the overdamped motion in a static potential $V(x)$ described by the Langevin equation \cite{horsthemke1984,risken1996fokker,kramers1940,hanggi1990}
\begin{equation}
  \frac{dx}{dt}=-\frac{dV(x)}{dx}+\sigma\xi(t),
  \label{eq:langevin}
\end{equation}
where $-V'(x)$ is the deterministic force and $\xi(t)$ is the Gaussian white noise satisfying $\langle \xi(t) \rangle=0$ and $\langle \xi(t)\xi(s) \rangle = \delta(t-s)$.
We assume $V(x)$ in a simple form of the polynomial
\begin{equation}
    V(x)=2-x+(x-2)^3.
    \label{eq:potential}
\end{equation}
The cubic profile (c.f., Fig.~\ref{fig:potential}) has two extrema located at $2\pm 1/\sqrt{3}$ and resembles the archetypal form of the Kramers potential \cite{kramers1940} or Michaelis-Menten chemical kinetics \cite{Marcinek}.

In the absence of resetting, the probability $p(x,t)=p(x,t|x_0,t_0)$ of finding the particle at time $t$ in the vicinity of $x$ evolves according to the (forward) Smoluchowski-Fokker-Planck equation \cite{goelrichter1974,gardiner2009}
\begin{equation}
    \frac{\partial p(x,t)}{\partial t} = \frac{\partial }{\partial x} \left[ V'(x)p(x,t) + \frac{\sigma^2}{2} \frac{\partial p(x,t)}{\partial x} \right].
    \label{eq:forward-fp}
\end{equation}

We explore the properties of noise-enhanced stability \cite{agudov2001,dubkov2004,fiasconaro2006co} under stochastic resetting \cite{evans2011diffusion,evans2011diffusion-jpa,evans2020stochastic}  as these two effects compete with each other.
More precisely, NES is responsible for elongating the decay of unstable states, while stochastic resetting is typically expected to decrease the lifetime and expedite the completion of the first passage process like, e.g., a chemical reaction.
To assess the interplay between these two stochastic effects, we explore the properties of the mean first passage time (MFPT) from the interval $(x_b,\infty)$, which is the average of the first passage times
\begin{equation}
    \mathcal{T} =   \langle t_{\mathrm{fp}} \rangle =
     \langle \min\{t : x(0)=x_0 \;\land\; x(t) < x_b  \} \rangle,
     \label{eq:mfpt-eq}
\end{equation}
where $x_b$ is the location of the absorbing boundary.
Therefore, the MFPT can be used to measure the lifetime, i.e., the average time needed to annihilate a particle at the absorbing boundary located at $x_b$.
As we will be further reinterpreting $x(t)$ as the concentration of reagents (or cells, see Sec.~\ref{sec:reinterpretation}),
by assumption $x_b$ cannot be negative. 
Hereafter, we set the barrier position to $x_b=0$.
For the uninterrupted motion, the MFPT designed as $\mathcal{T}(x)$ satisfies the equation \cite{goelrichter1974,gardiner2009}
\begin{equation}
    -V'(x) \frac{\partial \mathcal{T}(x)}{\partial x} + \frac{\sigma^2}{2} \frac{\partial^2 \mathcal{T}(x)}{\partial x^2}=-1
    \label{eq:backward-fp}
\end{equation}
with the additional condition $\mathcal{T}(x_b)=0$.

The action of  resetting \cite{evans2011diffusion,evans2011diffusion-jpa} is further addressed by assuming random restarts of the diffusive process occurring in time according to Poisson statistics.
The underlying diffusion process $\{X(t)\}$ starts at some point $x_0$ and evolves over a certain (random) time on the interval $(x_b,\infty)$ with deterministic forcing $-V'(x)$ powered by Gaussian white noise.
By a resetting mechanism, a particle diffusing in the potential $V(x)$ is revert to the point $x_0$ on the left hand side of the barrier $x_0 \in (0 , 2-1/\sqrt{3})$ and restarts its motion.

Without loss of generality, we can assume that the system starts (at $t=0$) its evolution at $x_0$, i.e., we study the system dynamics right after the first resetting.
The time duration $\tau$ between two subsequent resets is a random variable following the exponential distribution $\phi(\tau)=r \exp{(-r \tau)}$ with $r$ ($r>0$) denoting the resetting rate.
The effect of resetting can be assessed by examination of the coefficient of variation (CV) \cite{pal2017first}
\begin{equation}
    CV= \frac{\sigma(  t_{\mathrm{fp}})  }{ \langle t_{\mathrm{fp}} \rangle} = \frac{\sigma(  t_{\mathrm{fp}})}{\mathcal{T}},
    \label{eq:cv}
\end{equation}
which is the ratio between the standard deviation $\sigma(t_{\mathrm{fp}})$ of the first passage times and the mean first passage time $\mathcal{T}$ in the absence of stochastic resetting \cite{reuveni2016optimal,pal2019first}.
As explained in \cite{reuveni2016optimal,ray2021resetting} $CV$ can be used to find the domain where resetting can facilitate the escape kinetics.
Such a domain corresponds to $CV>1$ and it is the domain which we are interested in.
Noise enhances stability is observed in unstable potentials where the action of noise can increase the lifetime of unstable states.
The facilitation of the escape kinetics due to resetting can counteract the action of the noise-enhanced stability and make the elimination of tumor cells more efficient.


%
%
\section{Results\label{sec:results}}

We start with a detailed discussion of the ordinary noise-enhanced stability effect under stochastic resetting (Sec.~\ref{sec:nes}).
Afterwards, we reinterpret the model in the context of tumor dynamics (Sec.~\ref{sec:reinterpretation}) .

\subsection{Ordinary NES\label{sec:nes}}

The model described by Eqs.~(\ref{eq:langevin}) and~(\ref{eq:potential}) is studied by the method of stochastic dynamics.
Namely, the Euler-Maruyama method \cite{higham2001algorithmic,mannella2002} is applied. Therefore, $x(t+\Delta t)$ is approximated as
\begin{equation}
    x(t+\Delta t) = x(t) -V'(x(t)) \Delta t + \sigma \xi_i \sqrt{\Delta t},
\end{equation}
where $\xi_i$ is the sequence of independent identically distributed random variables following the standard normal distribution $N(0,1)$.

\begin{figure}[!h]
    \centering
    \includegraphics[angle=0,width=0.95\columnwidth]{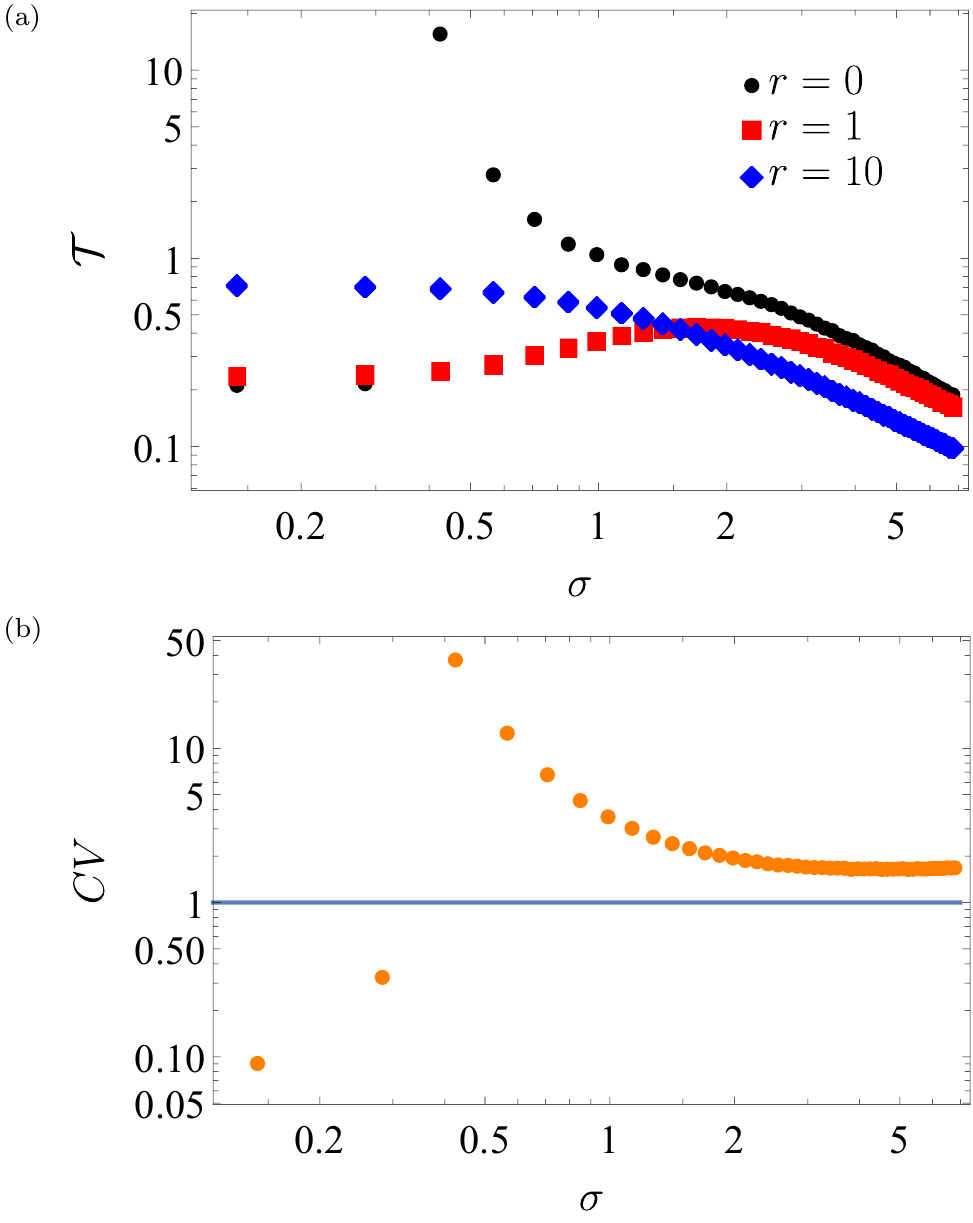}
    \caption{MFPT (top panel --- (a)) as a function of the noise intensity $\sigma$ for various values of resetting parameter $r$, together with corresponding values of coefficient of variation $CV$ (bottom panel --- (b)). The initial position is set to $x_0=1$.
    The absorbing boundary is located at $x_b=0$.
    }
    \label{fig:xreset}
\end{figure}

The top panel of Fig.~\ref{fig:xreset} presents the numerically estimated mean first passage time from $(x_b,\infty)$ with $x_b=0$ under the action of the deterministic force derived from the potential $V(x)$ given by Eq.~(\ref{eq:potential}) for various values of the resetting rate $r$.
$r=0$ corresponds to the case without resetting, i.e., to the noise-induced dynamics in an unstable potential, which can result in the occurrence of the noise-enhanced stability.
The absorbing boundary is located at $x_b=0$, while the initial position $x_0$ is set to $x_0=1$.
The deterministic force prevents the particle from escaping to infinity, therefore it effectively plays the role of the reflecting boundary, which is formally located at $+\infty$.
The bottom panel of Fig.~\ref{fig:xreset} displays the corresponding dependence of the coefficient of variation, see Eq.~(\ref{eq:cv}).

Examination of CV, see the bottom panel of Fig.~\ref{fig:xreset}, clearly indicates that in the majority of situations stochastic resetting  facilitates the escape kinetics.
The $CV$ becomes smaller than 1 only at a very low noise intensity $\sigma$ when the motion is practically deterministic.
For these low values of the noise intensity ($\sigma \approx 0$), the stochastic resetting interrupts the deterministic sliding, thus enhancing the lifetime of deterministically unstable states.
Also, for $\sigma \approx 0$, the MFPT grows semi-exponentially with the increase of the resetting rate $r$, see App.~\ref{sec:weaknoise}.
The conclusions drawn from the analysis of the $CV$ behavior are corroborated by studies of the MFPT, see the top panel of Fig.~\ref{fig:xreset}.
It shows that despite the fact that stochastic resetting can suppress the sliding to the absorbing boundary, it is typically not capable of enhancing the lifetime of unstable states when $\sigma \gg 0$.

Within our diffusion model incorporating NES and stochastic resetting, two distinct resetting regimes are visible. For $r\to 0$ resets are very rare, and the stochastic motion tends to the overdamped diffusion without restarts.
In this regime, starting motion anew increases the MFPT only for small noise intensities, as resetting hinders sliding towards the absorbing boundary, see App.~\ref{sec:weaknoise}.
In the opposite limit of $r\to\infty$, resets are so frequent that the particle practically cannot move freely because it is immediately restarted.
These frequent restarts are responsible for the diverging lifetimes of unstable states.
Nevertheless, such enhancement does not rely on any nontrivial interplay between resetting and stochastic dynamics as it simply ``sticks'' a particle to the initial position $x_0$.

\begin{figure}[!h]
    \centering
     \includegraphics[angle=0,width=0.95\columnwidth]{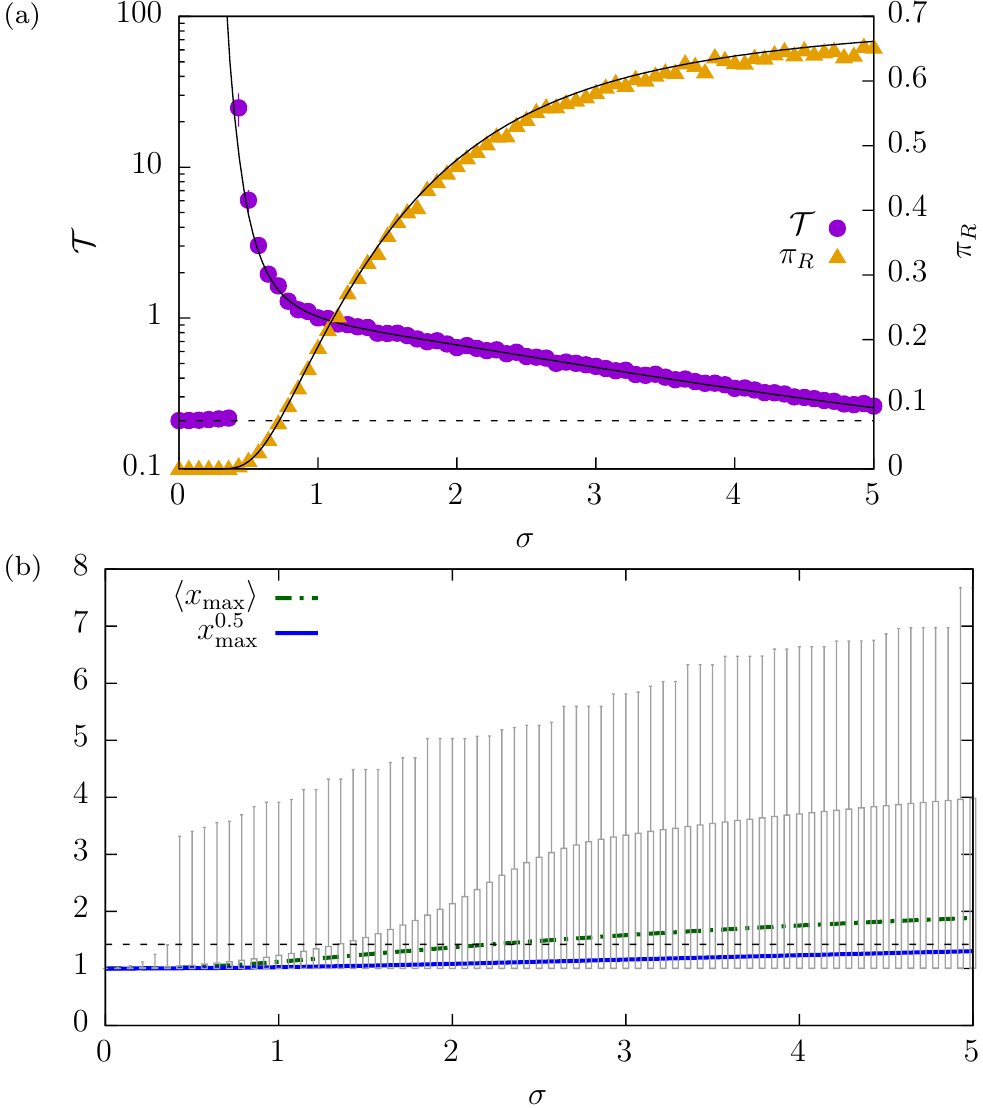}
    \caption{Top panel (a) presents MFPT as a function of the noise intensity $\sigma$ for  $r=0$ (left axis), together with corresponding values of the splitting probability $\pi_R$ (right axis).
    Solid lines show theoretical values of the MFPT, see Eq.~(\ref{eq:mfpt-ra-pot}), and splitting probability, see Eq.~(\ref{eq:splitting}).
    The dashed line displays the time of deterministic sliding $\mathcal{T}_{\sigma=0} \approx 0.21$.
    The bottom panel (b) shows $\langle x_{\mathrm{max}} \rangle$, median of $x_{\mathrm{max}}$ with quantiles of order 0.1 and 0.9 (boxes) along with minimal ($\min(\{x_{\mathrm{max}}\})$) and maximal ($\max(\{x_{\mathrm{max}}\})$) values of $x_{\mathrm{max}}$ (whiskers).
    The dashed line indicates the position of the top of the potential barrier, i.e., $x=2-1/\sqrt{3} \approx 1.42$.
    Initial position is set to $x_0=1$. The absorbing boundary is located at $x_b=0$.
    }
    \label{fig:nes}
\end{figure}

To find the reason why stochastic resetting shortens the lifetime of unstable states, we return to the examination of the classical NES effect in the potential given by Eq.~(\ref{eq:potential}).
For the resetting-free case, the MFPT can be calculated as the mean first passage time from the interval restricted by the reflecting boundary (located at $+\infty$) and the absorbing boundary placed at $x_b=0$ \cite{gardiner2009}.
The MFPT satisfying Eq.~(\ref{eq:mfpt-eq}) reads
\begin{equation}
\mathcal{T}(x)=\frac{2}{\sigma^2} \int_{x_b}^{x} \frac{d y}{\psi_-(y)} \int_{y}^{\infty}  \psi_-(z) d z,
\label{eq:mfpt-ra}
\end{equation}
where $\psi_\mp(x)$ is given by
\begin{equation}
    \psi_\mp(x) = \exp\left[ \mp \frac{2 V(x)}{\sigma^2} \right].
\end{equation}
From Eq.~(\ref{eq:mfpt-ra}) one gets the classical formula for the MFPT
\begin{equation}
    \mathcal{T}(x)=\frac{2}{\sigma^2} \int_{x_b}^{x} \exp\left[ \frac{2V(y)}{\sigma^2} \right]{d y} \int_{y}^{\infty}  \exp\left[ -\frac{2V(z)}{\sigma^2} \right]d z.
    \label{eq:mfpt-ra-pot}
\end{equation}
From the top panel of Fig.~\ref{fig:nes} it is visible that the numerically estimated MFPT perfectly follows the theoretical curve.
Moreover, in the limit of $\sigma\to 0$ the MFPT is given by
\begin{equation}
    \mathcal{T}=-\int_{x_0}^{x_b}\frac{1}{V'(y)}dy,
    \label{eq:mfpt-det}
\end{equation}
which is the time of the deterministic sliding from $x_0$ to $x_b$.
For $\sigma=0$ the MFPT can be calculated directly from the Langevin equation, see Eqs.~(\ref{eq:langevin}) and (\ref{eq:mfpt-det}).
In the very same regime, the MFPT can be also calculated from Eq.~(\ref{eq:backward-fp}).
The dependence of the MFPT on the noise intensity $\sigma$ is clearly non-monotonic, see the top panel of Fig.~\ref{fig:nes}.
There exists an optimal value of the $\sigma$ leading to the maximal lifetime.
Typically, the rapid increase in the MFPT is attributed to trajectories which manage to surmount the potential and overpass the potential barrier.
This qualitative observation can be analyzed in a more systematic way.

The MFPT does not fully characterize the escape kinetics.
Consequently, the examination of other measures can be insightful.
For instance, the escape process can be further characterized by the splitting probability.
For the motion in the finite interval $(a,b)$ ($a<b$) one can calculate the probability of leaving the interval via the particular boundary $a$ or $b$.
The probability $\pi_R(x)=\pi_b(x)$ of leaving $(a,b)$ to the right satisfies \cite{goelrichter1974,gardiner2009}
\begin{equation}
-V'(x) \frac{\partial \pi_b(x)}{\partial x}     +\frac{\sigma^2}{2} \frac{\partial^2 \pi_b(x)}{\partial x^2} = 0
\end{equation}
and it is given by
\begin{equation}
    \pi_b(x)=\frac{\int_a^x \psi_+(y) dy}{\int_a^b \psi_+(y) dy}=\frac{\int_a^x \exp\left[ \frac{2V(y)}{\sigma^2} \right] dy}{\int_a^b \exp\left[ \frac{2V(y)}{\sigma^2} \right] dy}.
    \label{eq:splitting}
    \end{equation}
The probability of escaping to the left ($\pi_L(x)$) can be calculated from the relation $\pi_R(x)=\pi_b(x)=1-\pi_a(x)=1-\pi_L(x)$.
The splitting probability can provide fruitful insights into the escape kinetics.
In the top panel of Fig.~\ref{fig:nes} using the right ordinate, in addition to MFPT, the probability of the first escape from $(0,2-1/\sqrt{3})$ to the right is depicted and compared with results of computer simulations (triangles) indicating perfect level of agreement.
In situations when $\pi_R>0$ there are some trajectories which prior to the absorption at $x_b=0$ reached the barrier top located at $x=2-1/\sqrt{3}$.
These trajectories have the possibility of sliding to the minimum of the potential and wander around $x \approx 2 + 1/\sqrt{3}$.
This in turn gives the chance of enhancing the lifetime by trapping the particle in the vicinity of the potential minima.
Comparison of the MFPT with the dependence of the $\pi_R$ indicates that MFPT grows significantly when $\pi_R$ becomes slightly larger than 0.
For instance, the maximal value of the MFPT in Fig.~\ref{fig:nes} is recorded for $\sigma\approx 0.42$ and it reads $\mathcal{T}=24.74$ with $\pi_R=0.004$, i.e, only $0.4$\% of trajectories overpassed the potential barrier prior to the absorption at $x_b=0$.
For larger noise intensities particles can not only more easily reach the top of the potential barrier but after overpassing the potential barrier they can more easily return back.
Consequently, for large values of $\sigma$ the increase in the MFPT is not as large as for small noise intensities and the MFPT is the decreasing function of $\sigma$, see Fig.~\ref{fig:nes}(a).

In order to fully resolve the origin of NES we explore the statistics of maximal $x_{\mathrm{max}}$ (most distant points to the right) visited by individual trajectories.
Bottom panel of Fig.~\ref{fig:nes} depicts $\langle x_{\mathrm{max}} \rangle$, median of $x_{\mathrm{max}}$ ($x_{\mathrm{max}}^{0.5}$), quantiles of order $0.1$ ($x_{\mathrm{max}}^{0.1}$) and $0.9$ ($x_{\mathrm{max}}^{0.9}$) along with minimal ($\min(\{x_{\mathrm{max}}\})$) and maximal ($\max(\{x_{\mathrm{max}}\})$) recorded values of $x_{\mathrm{max}}$.
Quantiles are presented as boxes, the bottom part of the box shows 0.1 quantile while top of the box 0.9 quantile.
Minimal value, which is practically the same as 0.1 quantile, and maximal values of $x_{\mathrm{max}}$ are depicted by whiskers.
Examination of $x_{\mathrm{max}}$ statistics, analogously like the splitting probability $\pi_R$, indicates that the initial increase in the MFPT is produced by a very few trajectories that managed to overpass the top of the barrier, i.e., to reach points located to the right of the potential barrier ($x>2-1/\sqrt{3}$), which position is denoted by the dashed line.
Please note that $x_{\mathrm{max}}^{0.9}$ rises above the boundary for $\sigma > 1.2$, i.e., well above the value of the noise intensity which maximizes the lifetime.
It means, that for $\sigma < 1.2$, not less than 90\% of most distant visited points are located to the left of the potential barrier.
Therefore, the increase in the MFPT is mainly determined by a very few trajectories which pass over the potential barrier and get trapped in the vicinity of the potential minima.
From the top panel of Fig.~\ref{fig:nes} it implies that the maximal MFPT is associated with $\pi_R=0.004$.
Therefore, only $0.4$\% of trajectories managed to overpass the potential barrier prior to the absorption at $x_b=0$.
These very few trajectories are responsible for the enormous increase in the MFPT at $\sigma \approx 0.42$.
Importantly, this observation provides justification why resetting typically decreases the lifetime of unstable states.
On the one hand it can suppress sliding to the absorbing boundary, but on the other it eliminates trajectories which manage to surmount the potential or to reach the potential minimum.
For $\sigma>0$ the MFPT is determined by trajectories which have accomplished climbing up the potential.
The gain due to surmounting can be eliminated by restarting the motion because it efficiently bounds the most distant visited point.
This is in line with results of \cite{ray2021resetting}, where Authors have shown that the interplay between the thermal and potential energy is the key factor determining the resetting efficiency.

\subsection{Model reinterpretation\label{sec:reinterpretation}}

The model defined by Eq.~(\ref{eq:langevin}) can be reinterpreted in the language of chemical kinetics \cite{kramers1940,cantisan2021stochastic,ray2021resetting,berera2019}.
Contrary to quantum tunneling process or nucleation, the escape problem in the Kramers approach is defined by analyzing a classical point particle escaping due to random forces. Contemporary applications of that scheme range from condensed matter and biological systems to high-energy physics and cosmological phase transitions \cite{berera2019}.

The kinetic scheme of tumor growth proposed in \cite{garay1978kinetic} involves combination of replication of transformed cells and immunological interaction of the host organism with transformed cells. Free effector cells like T-lymphocytes or killer cells form a complex with the transformed cells followed by lysis of tumor cells and dissociation of the complex to nonreplicating (or dead) tumor cells and free effector cells. The target population (concentration) of tumor cells evolves then \cite{fiasconaro2006co,li2021survival} according to the equation
\begin{equation}
   \frac{dx}{dt}=(1-\Theta x)x-\beta\frac{x}{x+1}+ \sigma\xi(t)
   \label{garay}
\end{equation}
resembling the Michaelis-Menten kinetics $X+Y\rightarrow E\rightarrow Y+P$. With $x$ representing the concentration of tumor cells, $y$ standing for concentration of effector cells and parameters $\beta=1.48$, $\Theta=0.25$, this model reflects bistability with minima $x=0$ and $x\approx2.08$ separated by a potential barrier at $x\approx 0.925$, thus capturing qualitatively features of the kinetic model defined by Eqs.~(\ref{eq:langevin}) -- (\ref{eq:potential}). In particular, the potential $V(x)$ derived for this model has an archetypal Kramers form similar to the cubic potential depicted in Fig.~\ref{fig:potential}.
The concentration of transformed cells changes due to the interplay of deterministic kinetics,  environmental fluctuations (white noise term) and applied therapy (resetting).
Depending on the system state ($x$ value) the action of the deterministic forces can increase or decrease the concentration of tumor cells.
The minimum of the potential located to the right of the barrier corresponds to the metastable state characterized by the high tumor concentration.
To eliminate tumor cells, it is necessary to overpass the potential barrier and reach the $x=0$ concentration.
The chances of spontaneous (deterministic) reaching $x=0$ are minimal as fluctuations are the only forces which can induce surmounting of the potential barrier separating stationary states.
On the one hand, if the applied therapy can move the system out of the metastable state associated with the potential well to the point located to the left of the barrier, it increases the chances of reaching the state of negligible tumor concentration.
On the other hand, the system moved out of the metastable state can be sensitive to the effect of noise-enhanced stability, which weakens the role of applied therapy.
Let us assume that the applied therapy (resetting) cannot fully eliminate the tumor and its action brings the system to the $x_0$ state with a lower concentration of tumor cells.
Without loss of generality, we can also assume that the system starts (at $t=0$) its evolution at $x_0$, i.e., we analyze the tumor evolution right after the first initiation of the (random) treatment.
Fig.~\ref{fig:xreset} indicates that the randomly applied therapy typically decreases the time needed to reach the $x=0$ state facilitating the overall efficiency, because it can move the system out the domain of motion associated with the metastable fixed point, see Fig.~\ref{fig:potential}.
Therefore, it facilitates the transition over the potential barrier.
The only exception is observed for very small $\sigma$ when frequent resetting can trap the particle in the vicinity of $x_0$.
The examination of the MFPT curve, see Fig.~\ref{fig:xreset}, provides justification why resetting typically increases the efficiency of the applied therapy.

%
%
\section{Summary and conclusions \label{sec:summary}}

Noise-enhanced stability belongs to the class of noise-induced effects.
It demonstrates that the action of noise can increase the lifetime of unstable states.
At the same time, stochastic resetting is a protocol that can accelerate kinetics and decrease the mean first passage time \cite{ray2021resetting}.
Opposite roles played by these two effects call for understanding whether their simultaneous actions can counterbalance.
By using numerical simulations and phenomenological arguments, we have shown that stochastic resetting typically is not sufficient to increase the lifetime of unstable states.
The exceptions to this observation are detected only in the vanishing noise limit or in the limit of infinitely frequent resetting.
For vanishing noise, stochastic resetting prevents deterministic sliding to the absorbing boundary, whereas for the infinite resetting rate the resetting protocol traps the particle in the neighborhood of the restarting position.

The generic noise-enhanced stability setup can also be used to describe tumor dynamics.
In such a case, the particle position is interpreted as the concentration of the cancer cells.
Following this line of interpretation, the effect of the noise-enhanced stability can slow down tumor elimination.
However, with the help of stochastic resetting, interpreted as applied therapy, the eradication of tumor cells can be more efficient, even if the eradication (resetting to negligible concentration of transformed cells) is performed at random times.

%
%
\section*{Acknowledgments}

This research was supported in part by PLGrid Infrastructure and by the National Science Center (Poland) grant 2018/31/N/ST2/00598.
KC would like to thank Marta Capa{\l}a for sharing the medical perspective.

\section*{Data availability}
The data that support the findings of this study are available from the corresponding author (KC) upon reasonable request.

%
%
\appendix
\section{Weak noise limit\label{sec:weaknoise}}

For $\sigma\to 0$ and $x_0\in (0,2-1/\sqrt{3})$ the particle described by Eq.~(\ref{eq:langevin}) deterministically slides towards the absorbing boundary.
The time of deterministic sliding $T$ can be calculated from the backward Smoluchowski-Fokker-Planck equation with $\sigma=0$, see Eq.~(\ref{eq:backward-fp}), or from the noise-free Langevin equation
\begin{equation}
    \frac{dx}{dt}=-V'(x),
\end{equation}
with the initial conditions $x(0)=x_0$ and $x(T)=x_b$.
The time $T$ is given by Eq.~(\ref{eq:mfpt-det}) of the main text and it is the mean first passage time for the uninterrupted motion.
As is visible from Figs.~\ref{fig:xreset} and~\ref{fig:nes} the MFPT is close to $T$ also for very small $\sigma$.
Here, we determine how, in the small $\sigma$ limit, the lifetime changes with resetting.
We will show that the lifetime is fully determined by the time of the deterministic sliding $T$ and the reset rate $r$.
The distribution of time intervals $\tau$ between two consecutive resets follows the exponential density
\begin{equation}
    \phi(\tau)=re^{-r \tau}.
\end{equation}
Stochastic resetting can restart the motion if the interresetting time $\tau$ is smaller than the sliding time $T$, i.e.,
if $\tau < T$, with the probability
\begin{equation}
    p=\int_0^T\phi(t)dt=1-e^{-rT}.
\end{equation}
For $r>0$, the recorded number of resets $n$ ($n\in\{0,1,\dots\}$) follows the geometric distribution.
\begin{equation}
    p_r(n)=p^{n}(1-p)
\end{equation}
characterized by the mean value
\begin{equation}
    \langle n \rangle = \frac{p}{1-p}
\end{equation}
and the variance
\begin{equation}
    \sigma^2(n) = \frac{p}{(1-p)^2}.
\end{equation}
In the limit of $r\to\infty$ the probability $p$ tends to 1, making the average number of resets infinite, which in turn is responsible for the unrestricted increase of the MFPT.

\begin{figure}[!h]
    \centering
    \includegraphics[angle=0,width=0.95\columnwidth]{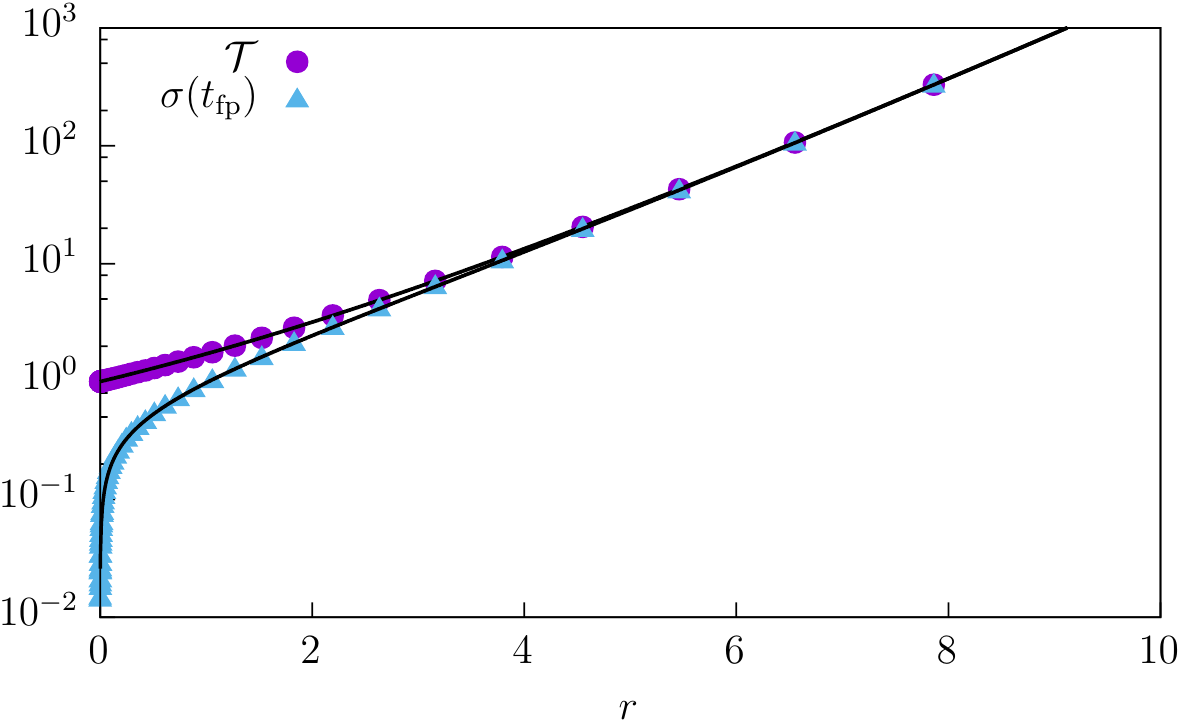}
    \caption{The average lifetime $\mathcal{T}$  (dots) and standard deviation $\sigma(t_{\mathrm{fp}})$ of first passage times (triangles) in the $\sigma\to 0$ limit.
    Points depict results of computer simulations, while solid lines are theoretical curves given by Eqs.~(\ref{eq:mfpt-s0-reset}) and~(\ref{eq:var-s0-reset}).
    }
    \label{fig:reset0sigma}
\end{figure}

The average lifetime $\mathcal{T}$ can be calculated as
\begin{equation}
    \mathcal{T}=\langle n \rangle \times \langle \tau | \tau < T\rangle +T,
    \label{eq:totaltime}
\end{equation}
where $\langle \tau | \tau < T\rangle$ is the conditional average of interresetting time following the exponential distribution restricted to $[0,T)$
\begin{equation}
    \langle \tau | \tau < T\rangle = \int_0^T  \frac{re^{-r\tau}}{1-e^{-rT}} \tau d\tau =  \frac{1}{r}-\frac{T}{e^{rT}-1}.
\end{equation}
The first term, $\langle n \rangle \times \langle \tau | \tau < T\rangle$, measures the total duration of slides interrupted by stochastic resets.
The last term in Eq.~(\ref{eq:totaltime}), $T$, quantifies the ultimate part of the motion, i.e, the time of sliding from $x_0$ to $x_b$.
Finally, we obtain the following formula for the MFPT under stochastic resetting
\begin{equation}
    \mathcal{T} = \langle t_{\mathrm{fp}} \rangle = \frac{e^{rT}-1}{r}.
    \label{eq:mfpt-s0-reset}
\end{equation}
The variance of the first passage time distribution can be calculated using the law of total variance
\begin{equation}
    \sigma^2(t_{\mathrm{fp}}) = \langle n \rangle \times \sigma^2( \tau | \tau < T ) + \sigma^2(n) \times \langle \tau | \tau < T \rangle^2,
\end{equation}
with
\begin{equation}
    \sigma^2( \tau | \tau < T ) = \frac{1}{r^2} + \frac{T^2}{2-2 \cosh(rT)}.
\end{equation}
Finally, one gets
\begin{equation}
    \sigma^2(t_{\mathrm{fp}}) = \frac{e^{2rT}-2e^{rT}r T-1}{r^2}=\frac{2e^{rT}\left[ \sinh(rT) -rT\right]}{r^2}.
    \label{eq:var-s0-reset}
\end{equation}
Both the mean lifetime and the variance of individual lifetimes grow semi-exponentially with $r$.
For $r=0$ the problem is fully deterministic with $\mathcal{T}$ given by Eq.~(\ref{eq:mfpt-det}).
In order to verify formulas~(\ref{eq:mfpt-s0-reset}) and~(\ref{eq:var-s0-reset}) we have performed simulations for $T$ arbitrarily, but without the loss of generality, set to $T=1$.
Fig.~\ref{fig:reset0sigma} presents the MFPT (circles) and the standard deviation of first passage times (triangles).
Solid lines correspond to theoretical curves, see Eqs.~(\ref{eq:mfpt-s0-reset}) and~(\ref{eq:var-s0-reset}).
Points, perfectly following theoretical predictions, represent the results of computer simulations.
The ratio of the theoretical and simulated values of MFPT and standard deviation is equal to unity with an accuracy not worse than 1\% (results not shown).
The stochastic resetting with a finite resetting rate $r$ is unable to fully eliminate the decay of unstable systems due to its stochastic character, i.e., the variability in the interresetting time.
Contrary to the Poissonian resetting, sharp resetting ($p(\tau)=\delta(\tau-\tau_0)$) is capable of producing infinite lifetime for finite (fixed) interresetting time $\tau_0$ ($\tau_0<T$).

%
%

%
%

\section*{References}
\def\url#1{}


\begin{thebibliography}{40}%
\makeatletter
\providecommand \@ifxundefined [1]{%
 \@ifx{#1\undefined}
}%
\providecommand \@ifnum [1]{%
 \ifnum #1\expandafter \@firstoftwo
 \else \expandafter \@secondoftwo
 \fi
}%
\providecommand \@ifx [1]{%
 \ifx #1\expandafter \@firstoftwo
 \else \expandafter \@secondoftwo
 \fi
}%
\providecommand \natexlab [1]{#1}%
\providecommand \enquote  [1]{``#1''}%
\providecommand \bibnamefont  [1]{#1}%
\providecommand \bibfnamefont [1]{#1}%
\providecommand \citenamefont [1]{#1}%
\providecommand \href@noop [0]{\@secondoftwo}%
\providecommand \href [0]{\begingroup \@sanitize@url \@href}%
\providecommand \@href[1]{\@@startlink{#1}\@@href}%
\providecommand \@@href[1]{\endgroup#1\@@endlink}%
\providecommand \@sanitize@url [0]{\catcode `\\12\catcode `\$12\catcode
  `\&12\catcode `\#12\catcode `\^12\catcode `\_12\catcode `\%12\relax}%
\providecommand \@@startlink[1]{}%
\providecommand \@@endlink[0]{}%
\providecommand \url  [0]{\begingroup\@sanitize@url \@url }%
\providecommand \@url [1]{\endgroup\@href {#1}{\urlprefix }}%
\providecommand \urlprefix  [0]{URL }%
\providecommand \Eprint [0]{\href }%
\providecommand \doibase [0]{http://dx.doi.org/}%
\providecommand \selectlanguage [0]{\@gobble}%
\providecommand \bibinfo  [0]{\@secondoftwo}%
\providecommand \bibfield  [0]{\@secondoftwo}%
\providecommand \translation [1]{[#1]}%
\providecommand \BibitemOpen [0]{}%
\providecommand \bibitemStop [0]{}%
\providecommand \bibitemNoStop [0]{.\EOS\space}%
\providecommand \EOS [0]{\spacefactor3000\relax}%
\providecommand \BibitemShut  [1]{\csname bibitem#1\endcsname}%
\let\auto@bib@innerbib\@empty
\bibitem [{\citenamefont {H\"anggi}, \citenamefont {Talkner},\ and\
  \citenamefont {Borkovec}(1990)}]{hanggi1990}%
  \BibitemOpen
  \bibfield  {author} {\bibinfo {author} {\bibfnamefont {P.}~\bibnamefont
  {H\"anggi}}, \bibinfo {author} {\bibfnamefont {P.}~\bibnamefont {Talkner}}, \
  and\ \bibinfo {author} {\bibfnamefont {M.}~\bibnamefont {Borkovec}},\
  }\bibfield  {title} {\enquote {\bibinfo {title} {Reaction-rate theory: Fifty
  years after {Kramers}},}\ }\href@noop {} {\bibfield  {journal} {\bibinfo
  {journal} {Rev. Mod. Phys.}\ }\textbf {\bibinfo {volume} {62}},\ \bibinfo
  {pages} {251} (\bibinfo {year} {1990})}\BibitemShut {NoStop}%
\bibitem [{\citenamefont {Anishchenko}\ \emph {et~al.}(1999)\citenamefont
  {Anishchenko}, \citenamefont {Neiman}, \citenamefont {Moss},\ and\
  \citenamefont {Schimansky-Geier}}]{anishchenko1999}%
  \BibitemOpen
  \bibfield  {author} {\bibinfo {author} {\bibfnamefont {V.~S.}\ \bibnamefont
  {Anishchenko}}, \bibinfo {author} {\bibfnamefont {A.~B.}\ \bibnamefont
  {Neiman}}, \bibinfo {author} {\bibfnamefont {F.}~\bibnamefont {Moss}}, \ and\
  \bibinfo {author} {\bibfnamefont {L.}~\bibnamefont {Schimansky-Geier}},\
  }\bibfield  {title} {\enquote {\bibinfo {title} {Stochastic resonance:
  Noise-enhanced order},}\ }\href@noop {} {\bibfield  {journal} {\bibinfo
  {journal} {Sov. Phys. Usp.}\ }\textbf {\bibinfo {volume} {42}},\ \bibinfo
  {pages} {7--36} (\bibinfo {year} {1999})}\BibitemShut {NoStop}%
\bibitem [{\citenamefont {Gammaitoni}\ \emph {et~al.}(2009)\citenamefont
  {Gammaitoni}, \citenamefont {H{\"a}nggi}, \citenamefont {Jung},\ and\
  \citenamefont {Marchesoni}}]{gammaitoni2009}%
  \BibitemOpen
  \bibfield  {author} {\bibinfo {author} {\bibfnamefont {L.}~\bibnamefont
  {Gammaitoni}}, \bibinfo {author} {\bibfnamefont {P.}~\bibnamefont
  {H{\"a}nggi}}, \bibinfo {author} {\bibfnamefont {P.}~\bibnamefont {Jung}}, \
  and\ \bibinfo {author} {\bibfnamefont {F.}~\bibnamefont {Marchesoni}},\
  }\bibfield  {title} {\enquote {\bibinfo {title} {Stochastic resonance: A
  remarkable idea that changed our perception of noise},}\ }\href@noop {}
  {\bibfield  {journal} {\bibinfo  {journal} {Eur. Phys. J. B}\ }\textbf
  {\bibinfo {volume} {69}},\ \bibinfo {pages} {1--3} (\bibinfo {year}
  {2009})}\BibitemShut {NoStop}%
\bibitem [{\citenamefont {McNamara}\ and\ \citenamefont
  {Wiesenfeld}(1989)}]{mcnamara1989}%
  \BibitemOpen
  \bibfield  {author} {\bibinfo {author} {\bibfnamefont {B.}~\bibnamefont
  {McNamara}}\ and\ \bibinfo {author} {\bibfnamefont {K.}~\bibnamefont
  {Wiesenfeld}},\ }\bibfield  {title} {\enquote {\bibinfo {title} {Theory of
  stochastic resonance},}\ }\href@noop {} {\bibfield  {journal} {\bibinfo
  {journal} {Phys. Rev. A}\ }\textbf {\bibinfo {volume} {39}},\ \bibinfo
  {pages} {4854--4869} (\bibinfo {year} {1989})}\BibitemShut {NoStop}%
\bibitem [{\citenamefont {Gammaitoni}\ \emph {et~al.}(1998)\citenamefont
  {Gammaitoni}, \citenamefont {H\"anggi}, \citenamefont {Jung},\ and\
  \citenamefont {Marchesoni}}]{gammaitoni1998}%
  \BibitemOpen
  \bibfield  {author} {\bibinfo {author} {\bibfnamefont {L.}~\bibnamefont
  {Gammaitoni}}, \bibinfo {author} {\bibfnamefont {P.}~\bibnamefont
  {H\"anggi}}, \bibinfo {author} {\bibfnamefont {P.}~\bibnamefont {Jung}}, \
  and\ \bibinfo {author} {\bibfnamefont {F.}~\bibnamefont {Marchesoni}},\
  }\bibfield  {title} {\enquote {\bibinfo {title} {Stochastic resonance},}\
  }\href@noop {} {\bibfield  {journal} {\bibinfo  {journal} {Rev. Mod. Phys.}\
  }\textbf {\bibinfo {volume} {70}},\ \bibinfo {pages} {223--287} (\bibinfo
  {year} {1998})}\BibitemShut {NoStop}%
\bibitem [{\citenamefont {Devoret}\ \emph {et~al.}(1984)\citenamefont
  {Devoret}, \citenamefont {Martinis}, \citenamefont {Esteve},\ and\
  \citenamefont {Clarke}}]{devoret1984}%
  \BibitemOpen
  \bibfield  {author} {\bibinfo {author} {\bibfnamefont {M.~H.}\ \bibnamefont
  {Devoret}}, \bibinfo {author} {\bibfnamefont {J.~M.}\ \bibnamefont
  {Martinis}}, \bibinfo {author} {\bibfnamefont {D.}~\bibnamefont {Esteve}}, \
  and\ \bibinfo {author} {\bibfnamefont {J.}~\bibnamefont {Clarke}},\
  }\bibfield  {title} {\enquote {\bibinfo {title} {Resonant activation from the
  zero-voltage state of a current-biased josephson junction},}\ }\href
  {\doibase 10.1103/PhysRevLett.53.1260} {\bibfield  {journal} {\bibinfo
  {journal} {Phys. Rev. Lett.}\ }\textbf {\bibinfo {volume} {53}},\ \bibinfo
  {pages} {1260--1263} (\bibinfo {year} {1984})}\BibitemShut {NoStop}%
\bibitem [{\citenamefont {Doering}\ and\ \citenamefont
  {Gadoua}(1992)}]{doering1992}%
  \BibitemOpen
  \bibfield  {author} {\bibinfo {author} {\bibfnamefont {C.~R.}\ \bibnamefont
  {Doering}}\ and\ \bibinfo {author} {\bibfnamefont {J.~C.}\ \bibnamefont
  {Gadoua}},\ }\bibfield  {title} {\enquote {\bibinfo {title} {Resonant
  activation over a fluctuating barrier},}\ }\href@noop {} {\bibfield
  {journal} {\bibinfo  {journal} {Phys. Rev. Lett.}\ }\textbf {\bibinfo
  {volume} {69}},\ \bibinfo {pages} {2318--2321} (\bibinfo {year}
  {1992})}\BibitemShut {NoStop}%
\bibitem [{\citenamefont {Magnasco}(1993)}]{magnasco1993}%
  \BibitemOpen
  \bibfield  {author} {\bibinfo {author} {\bibfnamefont {M.~O.}\ \bibnamefont
  {Magnasco}},\ }\bibfield  {title} {\enquote {\bibinfo {title} {Forced thermal
  ratchets},}\ }\href@noop {} {\bibfield  {journal} {\bibinfo  {journal} {Phys.
  Rev. Lett.}\ }\textbf {\bibinfo {volume} {71}},\ \bibinfo {pages}
  {1477--1481} (\bibinfo {year} {1993})}\BibitemShut {NoStop}%
\bibitem [{\citenamefont {Reimann}(2002)}]{reimann2002}%
  \BibitemOpen
  \bibfield  {author} {\bibinfo {author} {\bibfnamefont {P.}~\bibnamefont
  {Reimann}},\ }\bibfield  {title} {\enquote {\bibinfo {title} {Brownian
  motors: Noisy transport far from equilibrium},}\ }\href@noop {} {\bibfield
  {journal} {\bibinfo  {journal} {Phys. Rep.}\ }\textbf {\bibinfo {volume}
  {361}},\ \bibinfo {pages} {57--265} (\bibinfo {year} {2002})}\BibitemShut
  {NoStop}%
\bibitem [{\citenamefont {Russell}, \citenamefont {Wilkens},\ and\
  \citenamefont {Moss}(1999)}]{russell1999use}%
  \BibitemOpen
  \bibfield  {author} {\bibinfo {author} {\bibfnamefont {D.~F.}\ \bibnamefont
  {Russell}}, \bibinfo {author} {\bibfnamefont {L.~A.}\ \bibnamefont
  {Wilkens}}, \ and\ \bibinfo {author} {\bibfnamefont {F.}~\bibnamefont
  {Moss}},\ }\bibfield  {title} {\enquote {\bibinfo {title} {Use of behavioural
  stochastic resonance by paddle fish for feeding},}\ }\href@noop {} {\bibfield
   {journal} {\bibinfo  {journal} {Nature}\ }\textbf {\bibinfo {volume}
  {402}},\ \bibinfo {pages} {291--294} (\bibinfo {year} {1999})}\BibitemShut
  {NoStop}%
\bibitem [{\citenamefont {Simonotto}\ \emph {et~al.}(1997)\citenamefont
  {Simonotto}, \citenamefont {Riani}, \citenamefont {Seife}, \citenamefont
  {Roberts}, \citenamefont {Twitty},\ and\ \citenamefont
  {Moss}}]{simonotto1997visual}%
  \BibitemOpen
  \bibfield  {author} {\bibinfo {author} {\bibfnamefont {E.}~\bibnamefont
  {Simonotto}}, \bibinfo {author} {\bibfnamefont {M.}~\bibnamefont {Riani}},
  \bibinfo {author} {\bibfnamefont {C.}~\bibnamefont {Seife}}, \bibinfo
  {author} {\bibfnamefont {M.}~\bibnamefont {Roberts}}, \bibinfo {author}
  {\bibfnamefont {J.}~\bibnamefont {Twitty}}, \ and\ \bibinfo {author}
  {\bibfnamefont {F.}~\bibnamefont {Moss}},\ }\bibfield  {title} {\enquote
  {\bibinfo {title} {Visual perception of stochastic resonance},}\ }\href@noop
  {} {\bibfield  {journal} {\bibinfo  {journal} {Phys. Rev. Lett.}\ }\textbf
  {\bibinfo {volume} {78}},\ \bibinfo {pages} {1186} (\bibinfo {year}
  {1997})}\BibitemShut {NoStop}%
\bibitem [{\citenamefont {Agudov}(1998)}]{agudov1998noise}%
  \BibitemOpen
  \bibfield  {author} {\bibinfo {author} {\bibfnamefont {N.}~\bibnamefont
  {Agudov}},\ }\bibfield  {title} {\enquote {\bibinfo {title} {Noise delayed
  decay of unstable states},}\ }\href@noop {} {\bibfield  {journal} {\bibinfo
  {journal} {Phys. Rev. E}\ }\textbf {\bibinfo {volume} {57}},\ \bibinfo
  {pages} {2618} (\bibinfo {year} {1998})}\BibitemShut {NoStop}%
\bibitem [{\citenamefont {Agudov}\ and\ \citenamefont
  {Malakhov}(1999)}]{agudov1999decay}%
  \BibitemOpen
  \bibfield  {author} {\bibinfo {author} {\bibfnamefont {N.}~\bibnamefont
  {Agudov}}\ and\ \bibinfo {author} {\bibfnamefont {A.}~\bibnamefont
  {Malakhov}},\ }\bibfield  {title} {\enquote {\bibinfo {title} {Decay of
  unstable equilibrium and nonequilibrium states with inverse probability
  current taken into account},}\ }\href@noop {} {\bibfield  {journal} {\bibinfo
   {journal} {Phys. Rev. E}\ }\textbf {\bibinfo {volume} {60}},\ \bibinfo
  {pages} {6333} (\bibinfo {year} {1999})}\BibitemShut {NoStop}%
\bibitem [{\citenamefont {Agudov}\ and\ \citenamefont
  {Spagnolo}(2001)}]{agudov2001}%
  \BibitemOpen
  \bibfield  {author} {\bibinfo {author} {\bibfnamefont {N.~V.}\ \bibnamefont
  {Agudov}}\ and\ \bibinfo {author} {\bibfnamefont {B.}~\bibnamefont
  {Spagnolo}},\ }\bibfield  {title} {\enquote {\bibinfo {title} {Noise-enhanced
  stability of periodically driven metastable states},}\ }\href@noop {}
  {\bibfield  {journal} {\bibinfo  {journal} {Phys. Rev. E}\ }\textbf {\bibinfo
  {volume} {64}},\ \bibinfo {pages} {035102--035106} (\bibinfo {year}
  {2001})}\BibitemShut {NoStop}%
\bibitem [{\citenamefont {Fiasconaro}, \citenamefont {Valenti},\ and\
  \citenamefont {Spagnolo}(2003)}]{fiasconaro2003}%
  \BibitemOpen
  \bibfield  {author} {\bibinfo {author} {\bibfnamefont {A.}~\bibnamefont
  {Fiasconaro}}, \bibinfo {author} {\bibfnamefont {D.}~\bibnamefont {Valenti}},
  \ and\ \bibinfo {author} {\bibfnamefont {B.}~\bibnamefont {Spagnolo}},\
  }\bibfield  {title} {\enquote {\bibinfo {title} {Role of the initial
  conditions on the enhancement of the escape time in static and fluctuating
  potentials},}\ }\href@noop {} {\bibfield  {journal} {\bibinfo  {journal}
  {Physica A}\ }\textbf {\bibinfo {volume} {325}},\ \bibinfo {pages} {136}
  (\bibinfo {year} {2003})}\BibitemShut {NoStop}%
\bibitem [{\citenamefont {Dubkov}\ \emph {et~al.}(2020)\citenamefont {Dubkov},
  \citenamefont {Dybiec}, \citenamefont {Spagnolo}, \citenamefont {Kharcheva},
  \citenamefont {Guarcello},\ and\ \citenamefont
  {Valenti}}]{dubkov2020statistics}%
  \BibitemOpen
  \bibfield  {author} {\bibinfo {author} {\bibfnamefont {A.~A.}\ \bibnamefont
  {Dubkov}}, \bibinfo {author} {\bibfnamefont {B.}~\bibnamefont {Dybiec}},
  \bibinfo {author} {\bibfnamefont {B.}~\bibnamefont {Spagnolo}}, \bibinfo
  {author} {\bibfnamefont {A.}~\bibnamefont {Kharcheva}}, \bibinfo {author}
  {\bibfnamefont {C.}~\bibnamefont {Guarcello}}, \ and\ \bibinfo {author}
  {\bibfnamefont {D.}~\bibnamefont {Valenti}},\ }\bibfield  {title} {\enquote
  {\bibinfo {title} {Statistics of residence time for l{\'e}vy flights in
  unstable parabolic potentials},}\ }\href@noop {} {\bibfield  {journal}
  {\bibinfo  {journal} {Phys. Rev. E}\ }\textbf {\bibinfo {volume} {102}},\
  \bibinfo {pages} {042142} (\bibinfo {year} {2020})}\BibitemShut {NoStop}%
\bibitem [{\citenamefont {Pan}\ \emph {et~al.}(2009)\citenamefont {Pan},
  \citenamefont {Tan}, \citenamefont {Yu}, \citenamefont {Sun}, \citenamefont
  {Kang}, \citenamefont {Xu}, \citenamefont {Chen},\ and\ \citenamefont
  {Wu}}]{Pan2009}%
  \BibitemOpen
  \bibfield  {author} {\bibinfo {author} {\bibfnamefont {C.}~\bibnamefont
  {Pan}}, \bibinfo {author} {\bibfnamefont {X.}~\bibnamefont {Tan}}, \bibinfo
  {author} {\bibfnamefont {Y.}~\bibnamefont {Yu}}, \bibinfo {author}
  {\bibfnamefont {G.}~\bibnamefont {Sun}}, \bibinfo {author} {\bibfnamefont
  {L.}~\bibnamefont {Kang}}, \bibinfo {author} {\bibfnamefont {W.}~\bibnamefont
  {Xu}}, \bibinfo {author} {\bibfnamefont {J.}~\bibnamefont {Chen}}, \ and\
  \bibinfo {author} {\bibfnamefont {P.}~\bibnamefont {Wu}},\ }\bibfield
  {title} {\enquote {\bibinfo {title} {Resonant activation through effective
  temperature oscillation in a josephson tunnel junction},}\ }\href {\doibase
  10.1103/PhysRevE.79.030104} {\bibfield  {journal} {\bibinfo  {journal} {Phys.
  Rev. E}\ }\textbf {\bibinfo {volume} {79}},\ \bibinfo {pages} {030104}
  (\bibinfo {year} {2009})}\BibitemShut {NoStop}%
\bibitem [{\citenamefont {Yoshimoto}, \citenamefont {Shiraham},\ and\
  \citenamefont {Kurosawa}(2008)}]{Yoshi2008}%
  \BibitemOpen
  \bibfield  {author} {\bibinfo {author} {\bibfnamefont {M.}~\bibnamefont
  {Yoshimoto}}, \bibinfo {author} {\bibfnamefont {H.}~\bibnamefont {Shiraham}},
  \ and\ \bibinfo {author} {\bibfnamefont {S.}~\bibnamefont {Kurosawa}},\
  }\bibfield  {title} {\enquote {\bibinfo {title} {Noise-induced order in the
  chaos of the belousov-zhabotinsky reaction},}\ }\href@noop {} {\bibfield
  {journal} {\bibinfo  {journal} {J. Chem. Phys}\ }\textbf {\bibinfo {volume}
  {19}},\ \bibinfo {pages} {014508} (\bibinfo {year} {2008})}\BibitemShut
  {NoStop}%
\bibitem [{\citenamefont {Fiasconaro}\ \emph {et~al.}(2006)\citenamefont
  {Fiasconaro}, \citenamefont {Spagnolo}, \citenamefont {Ochab-Marcinek},\ and\
  \citenamefont {Gudowska-Nowak}}]{fiasconaro2006co}%
  \BibitemOpen
  \bibfield  {author} {\bibinfo {author} {\bibfnamefont {A.}~\bibnamefont
  {Fiasconaro}}, \bibinfo {author} {\bibfnamefont {B.}~\bibnamefont
  {Spagnolo}}, \bibinfo {author} {\bibfnamefont {A.}~\bibnamefont
  {Ochab-Marcinek}}, \ and\ \bibinfo {author} {\bibfnamefont {E.}~\bibnamefont
  {Gudowska-Nowak}},\ }\bibfield  {title} {\enquote {\bibinfo {title}
  {Co-occurrence of resonant activation and noise-enhanced stability in a model
  of cancer growth in the presence of immune response},}\ }\href@noop {}
  {\bibfield  {journal} {\bibinfo  {journal} {Phys. Rev. E}\ }\textbf {\bibinfo
  {volume} {74}},\ \bibinfo {pages} {041904} (\bibinfo {year}
  {2006})}\BibitemShut {NoStop}%
\bibitem [{\citenamefont {Zeng}\ \emph {et~al.}(2013)\citenamefont {Zeng},
  \citenamefont {Han}, \citenamefont {Yang}, \citenamefont {Wang},\ and\
  \citenamefont {Jia}}]{Zeng2013}%
  \BibitemOpen
  \bibfield  {author} {\bibinfo {author} {\bibfnamefont {C.}~\bibnamefont
  {Zeng}}, \bibinfo {author} {\bibfnamefont {Q.}~\bibnamefont {Han}}, \bibinfo
  {author} {\bibfnamefont {T.}~\bibnamefont {Yang}}, \bibinfo {author}
  {\bibfnamefont {H.}~\bibnamefont {Wang}}, \ and\ \bibinfo {author}
  {\bibfnamefont {Z.}~\bibnamefont {Jia}},\ }\bibfield  {title} {\enquote
  {\bibinfo {title} {Noise- and delay-induced regime shifts in an ecological
  system of vegetation},}\ }\href@noop {} {\bibfield  {journal} {\bibinfo
  {journal} {J. Stat. Mech. Theor. and Exp.}\ ,\ \bibinfo {pages} {P10017}}
  (\bibinfo {year} {2013})}\BibitemShut {NoStop}%
\bibitem [{\citenamefont {Fiasconaro}\ \emph {et~al.}(2008)\citenamefont
  {Fiasconaro}, \citenamefont {Ochab-Marcinek}, \citenamefont {Spagnolo},\ and\
  \citenamefont {Gudowska-Nowak}}]{Marcinek}%
  \BibitemOpen
  \bibfield  {author} {\bibinfo {author} {\bibfnamefont {A.}~\bibnamefont
  {Fiasconaro}}, \bibinfo {author} {\bibfnamefont {A.}~\bibnamefont
  {Ochab-Marcinek}}, \bibinfo {author} {\bibfnamefont {B.}~\bibnamefont
  {Spagnolo}}, \ and\ \bibinfo {author} {\bibfnamefont {E.}~\bibnamefont
  {Gudowska-Nowak}},\ }\bibfield  {title} {\enquote {\bibinfo {title}
  {Monitoring noise-resonant effects in cancer growth influenced by external
  fluctuations and periodic treatment},}\ }\href@noop {} {\bibfield  {journal}
  {\bibinfo  {journal} {Eur. Phys. J. B}\ }\textbf {\bibinfo {volume} {65}},\
  \bibinfo {pages} {435} (\bibinfo {year} {2008})}\BibitemShut {NoStop}%
\bibitem [{\citenamefont {Evans}\ and\ \citenamefont
  {Majumdar}(2011{\natexlab{a}})}]{evans2011diffusion}%
  \BibitemOpen
  \bibfield  {author} {\bibinfo {author} {\bibfnamefont {M.~R.}\ \bibnamefont
  {Evans}}\ and\ \bibinfo {author} {\bibfnamefont {S.~N.}\ \bibnamefont
  {Majumdar}},\ }\bibfield  {title} {\enquote {\bibinfo {title} {Diffusion with
  stochastic resetting},}\ }\href@noop {} {\bibfield  {journal} {\bibinfo
  {journal} {Phys Rev. Lett.}\ }\textbf {\bibinfo {volume} {106}},\ \bibinfo
  {pages} {160601} (\bibinfo {year} {2011}{\natexlab{a}})}\BibitemShut
  {NoStop}%
\bibitem [{\citenamefont {Evans}, \citenamefont {Majumdar},\ and\ \citenamefont
  {Schehr}(2020)}]{evans2020stochastic}%
  \BibitemOpen
  \bibfield  {author} {\bibinfo {author} {\bibfnamefont {M.~R.}\ \bibnamefont
  {Evans}}, \bibinfo {author} {\bibfnamefont {S.~N.}\ \bibnamefont {Majumdar}},
  \ and\ \bibinfo {author} {\bibfnamefont {G.}~\bibnamefont {Schehr}},\
  }\bibfield  {title} {\enquote {\bibinfo {title} {Stochastic resetting and
  applications},}\ }\href {\doibase 10.1088/1751-8121/ab7cfe} {\bibfield
  {journal} {\bibinfo  {journal} {J. Phys. A: Math. Theor.}\ }\textbf {\bibinfo
  {volume} {53}},\ \bibinfo {pages} {193001} (\bibinfo {year}
  {2020})}\BibitemShut {NoStop}%
\bibitem [{\citenamefont {Horsthemke}\ and\ \citenamefont
  {Lefever}(1984)}]{horsthemke1984}%
  \BibitemOpen
  \bibfield  {author} {\bibinfo {author} {\bibfnamefont {W.}~\bibnamefont
  {Horsthemke}}\ and\ \bibinfo {author} {\bibfnamefont {R.}~\bibnamefont
  {Lefever}},\ }\href@noop {} {\emph {\bibinfo {title} {Noise-inducted
  transitions. Theory and applications in physics, chemistry, and biology}}}\
  (\bibinfo  {publisher} {Springer Verlag},\ \bibinfo {address} {Berlin},\
  \bibinfo {year} {1984})\BibitemShut {NoStop}%
\bibitem [{\citenamefont {Risken}(9996)}]{risken1996fokker}%
  \BibitemOpen
  \bibfield  {author} {\bibinfo {author} {\bibfnamefont {H.}~\bibnamefont
  {Risken}},\ }\href@noop {} {\emph {\bibinfo {title} {The {Fokker-Planck}
  equation. Methods of solution and application}}}\ (\bibinfo  {publisher}
  {Springer Verlag},\ \bibinfo {address} {Berlin},\ \bibinfo {year}
  {19996})\BibitemShut {NoStop}%
\bibitem [{\citenamefont {Kramers}(1940)}]{kramers1940}%
  \BibitemOpen
  \bibfield  {author} {\bibinfo {author} {\bibfnamefont {H.~A.}\ \bibnamefont
  {Kramers}},\ }\bibfield  {title} {\enquote {\bibinfo {title} {Brownian motion
  in a field of force and the diffusion model of chemical reaction},}\
  }\href@noop {} {\bibfield  {journal} {\bibinfo  {journal} {Physica
  (Utrecht)}\ }\textbf {\bibinfo {volume} {7}},\ \bibinfo {pages} {284}
  (\bibinfo {year} {1940})}\BibitemShut {NoStop}%
\bibitem [{\citenamefont {Goel}\ and\ \citenamefont
  {Richter-Dyn}(1974)}]{goelrichter1974}%
  \BibitemOpen
  \bibfield  {author} {\bibinfo {author} {\bibfnamefont {N.~S.}\ \bibnamefont
  {Goel}}\ and\ \bibinfo {author} {\bibfnamefont {N.}~\bibnamefont
  {Richter-Dyn}},\ }\href@noop {} {\emph {\bibinfo {title} {Stochastic models
  in biology}}}\ (\bibinfo  {publisher} {Academic Press},\ \bibinfo {address}
  {New York},\ \bibinfo {year} {1974})\BibitemShut {NoStop}%
\bibitem [{\citenamefont {Gardiner}(2009)}]{gardiner2009}%
  \BibitemOpen
  \bibfield  {author} {\bibinfo {author} {\bibfnamefont {C.~W.}\ \bibnamefont
  {Gardiner}},\ }\href@noop {} {\emph {\bibinfo {title} {Handbook of stochastic
  methods for physics, chemistry and natural sciences}}}\ (\bibinfo
  {publisher} {Springer Verlag},\ \bibinfo {address} {Berlin},\ \bibinfo {year}
  {2009})\BibitemShut {NoStop}%
\bibitem [{\citenamefont {Dubkov}, \citenamefont {Agudov},\ and\ \citenamefont
  {Spagnolo}(2004)}]{dubkov2004}%
  \BibitemOpen
  \bibfield  {author} {\bibinfo {author} {\bibfnamefont {A.~A.}\ \bibnamefont
  {Dubkov}}, \bibinfo {author} {\bibfnamefont {N.~V.}\ \bibnamefont {Agudov}},
  \ and\ \bibinfo {author} {\bibfnamefont {B.}~\bibnamefont {Spagnolo}},\
  }\bibfield  {title} {\enquote {\bibinfo {title} {Noise-enhanced stability in
  fluctuating metastable states},}\ }\href@noop {} {\bibfield  {journal}
  {\bibinfo  {journal} {Phys. Rev. E}\ }\textbf {\bibinfo {volume} {69}},\
  \bibinfo {pages} {061103--061109} (\bibinfo {year} {2004})}\BibitemShut
  {NoStop}%
\bibitem [{\citenamefont {Evans}\ and\ \citenamefont
  {Majumdar}(2011{\natexlab{b}})}]{evans2011diffusion-jpa}%
  \BibitemOpen
  \bibfield  {author} {\bibinfo {author} {\bibfnamefont {M.~R.}\ \bibnamefont
  {Evans}}\ and\ \bibinfo {author} {\bibfnamefont {S.~N.}\ \bibnamefont
  {Majumdar}},\ }\bibfield  {title} {\enquote {\bibinfo {title} {Diffusion with
  optimal resetting},}\ }\href@noop {} {\bibfield  {journal} {\bibinfo
  {journal} {J. Phys. A: Math. Theor.}\ }\textbf {\bibinfo {volume} {44}},\
  \bibinfo {pages} {435001} (\bibinfo {year} {2011}{\natexlab{b}})}\BibitemShut
  {NoStop}%
\bibitem [{\citenamefont {Pal}\ and\ \citenamefont
  {Reuveni}(2017)}]{pal2017first}%
  \BibitemOpen
  \bibfield  {author} {\bibinfo {author} {\bibfnamefont {A.}~\bibnamefont
  {Pal}}\ and\ \bibinfo {author} {\bibfnamefont {S.}~\bibnamefont {Reuveni}},\
  }\bibfield  {title} {\enquote {\bibinfo {title} {First passage under
  restart},}\ }\href@noop {} {\bibfield  {journal} {\bibinfo  {journal} {Phys.
  Rev. Lett.}\ }\textbf {\bibinfo {volume} {118}},\ \bibinfo {pages} {030603}
  (\bibinfo {year} {2017})}\BibitemShut {NoStop}%
\bibitem [{\citenamefont {Reuveni}(2016)}]{reuveni2016optimal}%
  \BibitemOpen
  \bibfield  {author} {\bibinfo {author} {\bibfnamefont {S.}~\bibnamefont
  {Reuveni}},\ }\bibfield  {title} {\enquote {\bibinfo {title} {Optimal
  {{Stochastic Restart Renders Fluctuations}} in {{First Passage Times
  Universal}}},}\ }\href {\doibase 10.1103/PhysRevLett.116.170601} {\bibfield
  {journal} {\bibinfo  {journal} {Phys. Rev. Lett.}\ }\textbf {\bibinfo
  {volume} {116}},\ \bibinfo {pages} {170601} (\bibinfo {year}
  {2016})}\BibitemShut {NoStop}%
\bibitem [{\citenamefont {Pal}\ and\ \citenamefont
  {Prasad}(2019)}]{pal2019first}%
  \BibitemOpen
  \bibfield  {author} {\bibinfo {author} {\bibfnamefont {A.}~\bibnamefont
  {Pal}}\ and\ \bibinfo {author} {\bibfnamefont {V.~V.}\ \bibnamefont
  {Prasad}},\ }\bibfield  {title} {\enquote {\bibinfo {title} {First passage
  under stochastic resetting in an interval},}\ }\href {\doibase
  10.1103/PhysRevE.99.032123} {\bibfield  {journal} {\bibinfo  {journal} {Phys.
  Rev. E}\ }\textbf {\bibinfo {volume} {99}},\ \bibinfo {pages} {032123}
  (\bibinfo {year} {2019})}\BibitemShut {NoStop}%
\bibitem [{\citenamefont {Ray}\ and\ \citenamefont
  {Reuveni}(2021)}]{ray2021resetting}%
  \BibitemOpen
  \bibfield  {author} {\bibinfo {author} {\bibfnamefont {S.}~\bibnamefont
  {Ray}}\ and\ \bibinfo {author} {\bibfnamefont {S.}~\bibnamefont {Reuveni}},\
  }\bibfield  {title} {\enquote {\bibinfo {title} {Resetting transition is
  governed by an interplay between thermal and potential energy},}\ }\href@noop
  {} {\bibfield  {journal} {\bibinfo  {journal} {J. Chem. Phys.}\ }\textbf
  {\bibinfo {volume} {154}},\ \bibinfo {pages} {171103} (\bibinfo {year}
  {2021})}\BibitemShut {NoStop}%
\bibitem [{\citenamefont {Higham}(2001)}]{higham2001algorithmic}%
  \BibitemOpen
  \bibfield  {author} {\bibinfo {author} {\bibfnamefont {D.~J.}\ \bibnamefont
  {Higham}},\ }\bibfield  {title} {\enquote {\bibinfo {title} {An algorithmic
  introduction to numerical simulation of stochastic differential equations},}\
  }\href {\doibase 10.1137/S0036144500378302} {\bibfield  {journal} {\bibinfo
  {journal} {SIAM Review}\ }\textbf {\bibinfo {volume} {43}},\ \bibinfo {pages}
  {525--546} (\bibinfo {year} {2001})}\BibitemShut {NoStop}%
\bibitem [{\citenamefont {Mannella}(2002)}]{mannella2002}%
  \BibitemOpen
  \bibfield  {author} {\bibinfo {author} {\bibfnamefont {R.}~\bibnamefont
  {Mannella}},\ }\bibfield  {title} {\enquote {\bibinfo {title} {Integration of
  stochastic differential equations on a computer},}\ }\href@noop {} {\bibfield
   {journal} {\bibinfo  {journal} {Int. J. Mod. Phys. C}\ }\textbf {\bibinfo
  {volume} {13}},\ \bibinfo {pages} {1177--1194} (\bibinfo {year}
  {2002})}\BibitemShut {NoStop}%
\bibitem [{\citenamefont {Cantis{\'a}n}, \citenamefont {Seoane},\ and\
  \citenamefont {Sanju{\'a}n}(2021)}]{cantisan2021stochastic}%
  \BibitemOpen
  \bibfield  {author} {\bibinfo {author} {\bibfnamefont {J.}~\bibnamefont
  {Cantis{\'a}n}}, \bibinfo {author} {\bibfnamefont {J.~M.}\ \bibnamefont
  {Seoane}}, \ and\ \bibinfo {author} {\bibfnamefont {M.~A.}\ \bibnamefont
  {Sanju{\'a}n}},\ }\bibfield  {title} {\enquote {\bibinfo {title} {Stochastic
  resetting in the kramers problem: A monte carlo approach},}\ }\href@noop {}
  {\bibfield  {journal} {\bibinfo  {journal} {Chaos, Solitons \& Fractals}\
  }\textbf {\bibinfo {volume} {152}},\ \bibinfo {pages} {111342} (\bibinfo
  {year} {2021})}\BibitemShut {NoStop}%
\bibitem [{\citenamefont {Berera}\ \emph {et~al.}(2019)\citenamefont {Berera},
  \citenamefont {Mabillard}, \citenamefont {Mintz},\ and\ \citenamefont
  {Ramos}}]{berera2019}%
  \BibitemOpen
  \bibfield  {author} {\bibinfo {author} {\bibfnamefont {A.}~\bibnamefont
  {Berera}}, \bibinfo {author} {\bibfnamefont {J.}~\bibnamefont {Mabillard}},
  \bibinfo {author} {\bibfnamefont {B.}~\bibnamefont {Mintz}}, \ and\ \bibinfo
  {author} {\bibfnamefont {R.}~\bibnamefont {Ramos}},\ }\bibfield  {title}
  {\enquote {\bibinfo {title} {Formulating the kramers problem in field
  theory},}\ }\href@noop {} {\bibfield  {journal} {\bibinfo  {journal} {Phys.
  Rev. D}\ }\textbf {\bibinfo {volume} {100}},\ \bibinfo {pages} {076005}
  (\bibinfo {year} {2019})}\BibitemShut {NoStop}%
\bibitem [{\citenamefont {Garay}\ and\ \citenamefont
  {Lefever}(1978)}]{garay1978kinetic}%
  \BibitemOpen
  \bibfield  {author} {\bibinfo {author} {\bibfnamefont {R.~P.}\ \bibnamefont
  {Garay}}\ and\ \bibinfo {author} {\bibfnamefont {R.}~\bibnamefont
  {Lefever}},\ }\bibfield  {title} {\enquote {\bibinfo {title} {A kinetic
  approach to the immunology of cancer: Stationary states properties of
  efffector-target cell reactions},}\ }\href@noop {} {\bibfield  {journal}
  {\bibinfo  {journal} {J. Theor. Biol.}\ }\textbf {\bibinfo {volume} {73}},\
  \bibinfo {pages} {417--438} (\bibinfo {year} {1978})}\BibitemShut {NoStop}%
\bibitem [{\citenamefont {Li}\ \emph {et~al.}(2021)\citenamefont {Li},
  \citenamefont {Zhang}, \citenamefont {Yan},\ and\ \citenamefont
  {Xing}}]{li2021survival}%
  \BibitemOpen
  \bibfield  {author} {\bibinfo {author} {\bibfnamefont {D.}~\bibnamefont
  {Li}}, \bibinfo {author} {\bibfnamefont {N.}~\bibnamefont {Zhang}}, \bibinfo
  {author} {\bibfnamefont {M.}~\bibnamefont {Yan}}, \ and\ \bibinfo {author}
  {\bibfnamefont {Y.}~\bibnamefont {Xing}},\ }\bibfield  {title} {\enquote
  {\bibinfo {title} {Survival analysis for tumor growth model with stochastic
  perturbation},}\ }\href@noop {} {\bibfield  {journal} {\bibinfo  {journal}
  {Discrete Contin. Dyn. Syst. B}\ }\textbf {\bibinfo {volume} {26}},\ \bibinfo
  {pages} {5707} (\bibinfo {year} {2021})}\BibitemShut {NoStop}%
\end{thebibliography}
\end{document}